\documentclass[conference]{IEEEtran}
\usepackage[T1]{fontenc}
\usepackage{newtxtext}
\usepackage[utf8]{inputenc}
\usepackage{amsmath,amssymb}
\usepackage{textcomp}
\usepackage{xcolor}
\usepackage{array}
\usepackage{tabularx}
\usepackage{ragged2e}
\usepackage{listings}
\usepackage{xurl}
\usepackage[hidelinks]{hyperref}
\usepackage[capitalise]{cleveref}

\usepackage{graphicx}
\usepackage{booktabs}
\usepackage{footnote}

\newcolumntype{L}{>{\raggedright\arraybackslash}X}

\newcommand{\custompara}[1]{\vspace{0.2em}\textbf{\textit{#1.}}}

\definecolor{mycyan}{RGB}{51, 204, 255}
\definecolor{mypurple}{RGB}{153, 102, 255}
\definecolor{mypink}{RGB}{255, 51, 204}
\definecolor{myorange}{RGB}{255, 153, 102}
\definecolor{myteal}{RGB}{0, 204, 153}

\begin{document}

\title{Can Open Be Too Open? On the Security and Trustworthiness of Fully Transparent LLMs}

\author{\IEEEauthorblockN{Sheila Zingg}
\IEEEauthorblockA{ETH Zurich}
\and
\IEEEauthorblockN{Florian Tramèr}
\IEEEauthorblockA{ETH Zurich}
\and
\IEEEauthorblockN{Martin Vechev}
\IEEEauthorblockA{ETH Zurich}
\and
\IEEEauthorblockN{Srdjan Čapkun}
\IEEEauthorblockA{ETH Zurich}}

\maketitle

\begin{abstract}
There are different levels of transparency for LLMs: open weights, open data, and/or open pipeline. Recently, more interest has emerged in \textit{fully transparent LLMs}, i.e., LLMs with fully open weights, data, and training pipeline. Drawing analogies to other fields such as software security, it is often believed that increasing transparency inherently increases the security and trustworthiness of the LLM. However, we argue that such claims need to be critically re-evaluated in the context of LLMs. In this paper, we analyze the impact of transparency on LLM security and trustworthiness by reviewing the relevant literature. 
Our core findings are that, due to the limited reproducibility of LLM pipelines, in practice, users cannot verify the correspondence between open weights and the associated open data and pipeline. Thus, in order to trust a model, users still need to trust that developers acted honestly, reducing the benefits over models that provide less ``transparency''.
Moreover, we find that even when LLMs are fully transparent and trusted, this openness can favor adversaries. Open weights in particular, but also open data and pipelines, can increase the success rate of attacks. Although the same mechanisms can be used to improve the model and protect it against attacks, the current lack of robust lightweight mechanisms to patch LLM vulnerabilities limits the benefits of openness at this time. We note that our analysis is based on current LLM capabilities and analysis tools, and that future developments may require a re-evaluation of these points.
\end{abstract}

\section{Introduction}

In the current LLM landscape, there are different levels of model transparency (also called openness). Models can be made available with their weights (open-weight models), their data (open-data models), and/or their training pipeline (open-pipeline models). A model could have any combination of open/closed weights, open/closed data, and open/closed pipeline. However, most commonly used models are either fully closed, meaning that they are only accessible over an API and nothing is published (e.g., OpenAI GPT-5, Anthropic Claude Opus, Google Gemini 3), only open-weight (e.g., Meta Llama 4, Alibaba Qwen 3, Mistral Large 3), or fully open, meaning open weights, open data, and open pipeline (e.g., Ai2 OLMo, EleutherAI Pythia, Apertus 1.5, Lucie-7B). It must be noted that even for fully transparent LLMs, opaque components may have been used. For example, open data means that all data used for training is published; however, parts of the data are often generated by an LLM that may not be fully transparent, which is an opaque component. Increased transparency is often seen as improving security and trustworthiness~\cite{hall2025marin, llama_release_article, lawfare_open_models, linux_security_LLM_transparency, mozilla_joint_Statement}, similarly to how open-source software is often seen as providing better security guarantees and higher trustworthiness than closed-source software (i.e., the ``Many-Eyes Principle'' that states that given enough independent reviewers all bugs are shallow~\cite{raymond2001cathedral}). 

In this work, we argue, however, that these claims need to be evaluated carefully since the analogy between LLMs and software does not hold in the context of security and trustworthiness. Software code was designed to be read and edited by humans and can be deterministically and inexpensively compiled into a binary. In contrast, LLMs are less amenable to public scrutiny, even under full transparency, due to a combination of factors including the way in which the training data is selected, the non-determinism of training, the cost of training, and the lack of interpretability of the weights. Some may argue that transparency alone may lead to more honest development processes as even a slightly increased risk of detection could disincentivize malicious behavior. However, if unwarranted, this can also give a false sense of security, and thus, we believe it is important to identify the impact of transparency on the trust and security of LLMs.

To evaluate the security of an LLM, different adversaries must be considered: a) an inference-time adversary who attempts to use the model for malicious purposes, b) an external model poisoner who posts malicious data to frequently scraped websites or redistributes a malicious version of the model, c) a malicious insider who tries to inject weaknesses into a model by changing the training pipeline, training data, or weights, and d) a malicious model developer (i.e., the organization that develops the model) who ships or serves a different model than the one the user expects. These correspond to general classes of adversaries for software, namely a malicious user, a software supply-chain attacker, a malicious insider, and a malicious vendor. For software, open-sourcing can help against all of these adversaries; for LLMs, this is not the case.

\custompara{Transparency and Trust}
The discussion of trust in LLMs focuses primarily on adversary (d), i.e., whether a user can verify that the LLM they run has the advertised properties. Transparency is usually seen as a way to reduce the need for users to trust the model developer. In open-source software, the user can inspect the code and can verify that the program they are running resulted from the published code (e.g., through reproducible builds). Even more importantly, the user can verify that the implementation largely corresponds to the specification and the functionality that the software is supposed to implement. Given that there is a clear semantic link between the code and the program behavior, it is difficult for the developer to insert hidden functionality into the open-source code without eventually being detected by reviewers of the code. 
Thus, one can, with appropriate care (i.e., the software is indeed reviewed by many and is written and built in a way that can be understood and verified), rely on popular open-source codebases being trustworthy. 

For LLMs, this is usually not the case, even if a developer publishes the weights, data, and training pipeline of the model. The model weights do not provide a clear semantic link to the model's behavior, and the training data is too vast to inspect in sufficient depth, whether manually or with automated tools. A malicious developer can poison an LLM by inserting a backdoor in the data, which often requires only a few poisoned documents in the dataset, and can be hidden to make detection more difficult~\cite{carlini2024poisoning, souly2025poisoning, dragunsai, goldwasser2022planting}. This means that it would be hard for a user to identify a poisoned model even if they have access to the data and training pipeline. Furthermore, even if a malicious backdoor is found, it is hard to provably attribute it to the developer, since the data they collected could already have been poisoned by another party. Even if we assume that the user can be certain that the published training data does not contain any backdoors, they may not be able to trust the weights that they received. Current frontier training processes are non-deterministic (i.e., using the same training pipeline on the same training data twice will not result in the same weights)~\cite{rivera2021challenge, pham2020problems, zhuang2022randomness}, making it harder for users to verify that the weights actually resulted from applying the published training pipeline on the published data. Furthermore, even if training were deterministic, there are very few entities that could retrain a model to verify its correctness given the high effort required for training and the difficulty of replicating exact training hardware and entire training pipelines. Thus, unlike for open-source software, the user of a transparent LLM still needs to trust the developer to a significant degree. An overview of these  differences is shown in \cref{tab:trust-assumptions}.

\begin{table*}[t]
\centering
\caption{Developer trust assumptions for closed LLMs, fully transparent LLMs, and open-source software.}
\label{tab:trust-assumptions}
\footnotesize
\renewcommand{\arraystretch}{1.3}
\newcolumntype{L}{>{\raggedright\arraybackslash}X}
\begin{tabularx}{\textwidth}{>{\raggedright\arraybackslash\bfseries}p{0.13\textwidth} L L L}
\toprule
\textbf{Aspect} & \textbf{Closed LLM} & \textbf{Fully transparent LLM} & \textbf{Open-source software} \\
\midrule
Trust placed in the developer
 & The developer must be trusted for the full model development and provision. This includes that the API serves the advertised model, that the model was trained using a safe process, that no malicious capability was inserted, and that future updates do not change the behavior.
 & The developer needs to be trusted that the published weights resulted from using the published training pipeline on the published training data. Furthermore, the training data needs to be trusted to be free from concealed backdoors. Lastly, if non-transparent LLMs are used as part of the development process, then the developers of those models need to be partially trusted.
 & In principle, little trust in the developer is needed if the source code is written well, reviewed by many, and can be rebuilt and verified against the intended specification. In practice, residual trust remains in code completeness, dependencies, build tooling, signing infrastructure, and the adequacy of community review/testing. \\
\midrule
Practical verification methods available to a user
 & Users can only use black-box checks. These include prompt-response behavior, behavioral fingerprinting to detect model substitution, and independently reproduced benchmarks.
 & Users can verify the integrity of the artifacts, which include weight hashes/signatures, data-snapshot hashes, and checkpoint metadata. Furthermore, they can check the pipeline code, analyze the data (e.g., checking for benchmark data), perform white-box weight analysis (e.g., weight statistics, activation- or gradient-based probing), and partially retrain to approximately verify parts of the training. Lastly, they can use all black-box checks.
 & Users can verify the integrity of the artifacts, which include source code hashes, signatures, and reproducible builds linking the executable to the published source code. Furthermore, they can use code-property checks, which include code review, code testing (e.g., fuzzing), static and dynamic analysis, and formal protocol/program verification. \\
\midrule
Constraints on the adversary capabilities of a malicious developer
 & The adversary must preserve perceived model behavior, avoid obviously malicious outputs, evade user monitoring/complaints, and pass required safety evaluations or legally mandated audits.
 & The adversary must conceal backdoors or pipeline changes in the published data/code/weights so that they evade inspection, statistical checks, behavioral probes, and audits. All constraints from the closed LLM also apply.
 & The adversary must make malicious code appear benign to reviewers, evade tests and build scanning, and hide payloads in complex logic, toolchains, or build/release pipelines. \\
\bottomrule
\end{tabularx}
 
\vspace{0.5em}
\begin{minipage}{\textwidth}
\footnotesize
\textit{Note.} This table focuses on adversary (d). Verification refers to practical, resource-bounded checks available to an external user, not formal proofs.
\end{minipage}
\end{table*}

To resolve the limited reproducibility, proof-of-learning~\cite{jia2021proof}, zero-knowledge proofs of training~\cite{garg2023experimenting}, and verifiable training~\cite{srivastava2024optimistic} have been proposed as methods with which a user can efficiently verify that a model was trained using the published training pipeline on the published training data. However, proof-of-learning is not robust enough to provide sufficient guarantees~\cite{fang2023proof} and zero-knowledge proofs of training are too expensive for the model creator to be practical. Verifiable training has been used effectively for smaller 1B models~\cite{donaghy2026open}, but is not yet efficient enough for frontier models and places high resource requirements on the verifier. While this means that, currently, users of frontier models must trust a third party for LLM provenance, this third party may not necessarily have to be the model developer. For example, in attested training~\cite{duddu2024laminator, spoczynski2025atlas}, the relevant parts of the training pipeline are executed within a trusted execution environment (TEE) that was attested by another third party (typically the CPU manufacturer). Thus, the user can be certain that the LLM was trained as published as long as they trust the attesting party. However, attested training only shifts the trust and has only been shown viable for small models and model finetuning, primarily due to hardware limitations of TEEs. Once technical limitations are resolved, these techniques could be used to validate the model, in which case transparency could reduce the required trust in the developer, although trust in the attestation authority and in the robustness of TEEs would still be required.

\custompara{Transparency and Security}
If the model developer is trusted (i.e., considering adversaries (a)--(c)), transparency is often considered to enable a user to identify vulnerabilities before they are impacted by them. However, such arguments do not hold for LLMs directly. 
LLMs are inherently generative, and thus, even if we trust the developer and assume that the training data is fully free of backdoors, the LLM may still generate unsafe outputs. Thus, the LLM needs to be scanned for vulnerabilities continuously so that they can be resolved. 
However, when a vulnerability is identified, unlike with code, there is no simple and robust way in which LLMs can be patched. Lightweight methods like unlearning and safety finetuning are not robust~\cite{hubinger2024sleeper, lucki2024adversarial}, and output filtering cannot be enforced for local deployments, which are always possible when the weights are published. For robust patching, an LLM would likely need to be retrained, which is resource-intensive.
Furthermore, weight transparency allows adversaries to cost-effectively remove any safety alignment and guardrails~\cite{qi2024fine, arditi2024refusal, lermen2023lora}, meaning that hardening the model against attacks may not yield better security outcomes.
Lastly, transparency may even help an adversary who wants to exploit a transparent LLM, as they can use knowledge about the weights~\cite{schwinn2024soft, huang2024catastrophic} in particular, but also the data~\cite{deng2024multilingual, yong2023low, wang2024all, shen2024language} and pipeline~\cite{hackett2025bypassing, nasr2025attacker, wang2025protect} to improve their attacks.

\custompara{Cost of Transparency}
Although full model transparency can reduce inference costs, it can also introduce costs. Transparency can introduce legal risk for model developers~\cite{bartz2025anthropic, nyt2023microsoft, kadrey2025meta}, as LLMs are often trained on scrapes of the internet, which can contain sensitive data, such as copyrighted material or PII. For closed-data models, it is harder to prove that they were trained on sensitive data, while open-data models allow anyone to search for such data in the published dataset. While the exact legal basis is still unclear~\cite{bommasani2024considerations}, this increases the liability risk for developers of open-data models. There are methods to remove sensitive data from the dataset~\cite{lukas2023analyzing} or to reduce the probability that an LLM will repeat its training data verbatim~\cite{hans2024like}. However, these methods are imperfect, and it is unclear whether they will be considered sufficient in a court case. While this mainly impacts the model developer, it may also affect model users (e.g., by losing access to the model or through downstream liability). If an open-data LLM is used by an influential party (e.g., a government), it may be scrutinized more, increasing the risk of lawsuits. In fact, the legal uncertainty around training data has already resulted in a decline of open training data~\cite{bengio2025international, longpre2024large}.

These legal considerations combined with ethical challenges when publishing certain training data (e.g., human-annotated data) may make transparent model developers more cautious and restrictive with the training data that they use, which can impact model performance. 
For example, it has been shown that using high-quality proprietary data for training increases model performance~\cite{wu2023bloomberggpt, yang2022large, singhal2025toward}. However, data holders of proprietary data are unlikely to agree to publish this data. Some transparent model developers have adapted to this by publishing most instead of all data~\cite{soofi2026sovereign, blakeman2025nvidia}, allowing them to train on proprietary data without publishing it.

\custompara{Independence}
Another benefit of transparency is that access to the model cannot be removed or restricted after the model has been published as long as the user has access to sufficient compute to run the model locally. Furthermore, the user of the model can adapt the model's behavior to fit their needs. While this alone is not enough to provide full sovereignty, it does reduce the dependency on third parties.

\custompara{When to use a Transparent Model}
In summary, currently, there is no level of transparency that is always best for users. Instead, users should analyze different factors (e.g., the trustworthiness of the developer, what they want to use the model for, their security requirements, legal implications, and their technical capabilities) to decide what level of transparency would bring the most benefits while limiting risks. Furthermore, the two ends of the spectrum, fully closed and fully open models, should not be seen as the only ones worth exploring. It might be that intermediate levels of transparency (e.g., an open-weight, open-pipeline, closed-data model) may be the best-suited option for a specific use case. This means that each component of transparency (weights, data, and pipeline) needs to be considered individually when making an assessment of the risks and benefits of a model. Lastly, the way transparency is evaluated needs to be updated continuously as new tools and methods to analyze LLMs emerge. A component that is currently not helpful in establishing trust or increasing security might be able to do so in the future.

\custompara{Our Contributions}
In this paper, we investigate the impact of transparency on the security and trustworthiness of LLMs based on the current capabilities of LLMs and the currently available LLM analysis methods and tools:
\begin{itemize}
    \item \textbf{Discussion of the differences between open-source software and fully transparent LLMs:} We review why the claimed benefits from open-source software cannot be applied directly to fully transparent LLMs. These differences help guide discussions on security claims.
    \item \textbf{Overview of the impact of transparency on security and trust:} Across the LLM components, we discuss where transparency favors defenders and where it may favor attackers. We also discuss how transparency impacts trust and what methods and technologies may help.
    \item \textbf{Discussion of how users can evaluate transparency:} We list aspects that may be important when evaluating what level of transparency is desirable. We do not recommend a specific level of transparency and instead give an overview of what to consider when making a decision.
\end{itemize}
\section{Background}

\begin{figure*}[tb]
    \centering
    \includegraphics[clip, trim=1cm 23cm 1cm 1cm, width=\linewidth]{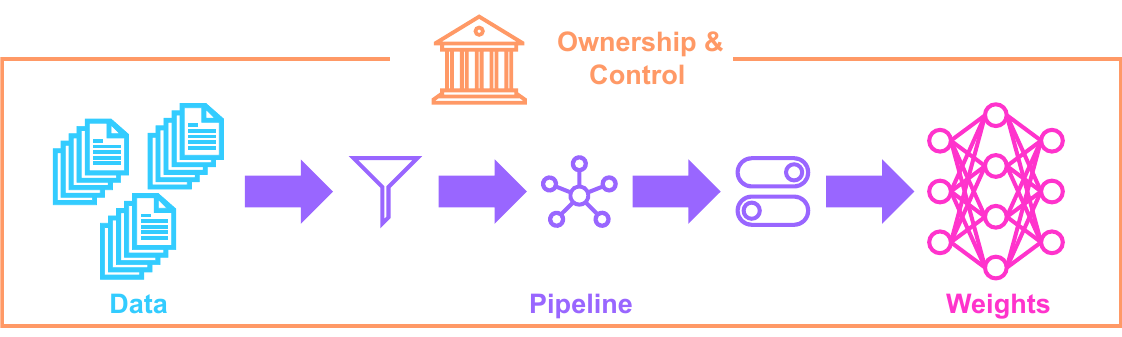}
    \caption{Overview of the components that are part of LLM training. In blue is the {\color{mycyan}training data}, in purple the {\color{mypurple}training pipeline}, in pink the {\color{mypink}model weights}, and in orange who {\color{myorange}owns and controls the model}.}
    \label{fig:llm_overview}
\end{figure*}

\cref{fig:llm_overview} shows the different components of LLM training, based on which we describe the terms that are often used when discussing transparency. This allows us to define the different degrees of LLM transparency, including full transparency. Then we discuss how the transparency of an LLM relates to its sovereignty. Lastly, we describe the different threat models that are relevant for LLM security.

\subsection{Terms}
\label{subsec:terms}

\custompara{\color{mycyan}Open Data} 
Open data means that the model developer publishes which data they used to train the model. This includes data from all training stages, such as pretraining, mid-training (e.g., domain- or capability-specific data such as code, mathematics, or long-context data), supervised finetuning (e.g., instruction, reasoning, and safety data), and preference or reinforcement learning (e.g., preference pairs, reward model training data, and RL environments). If parts of the data are synthetically generated (e.g., by another LLM), open data includes the generated data and information on how it was generated. It is possible for model developers to only publish the data partially. This is usually considered open data as long as it contains a large portion of the training data~\cite{liesenfeld2024rethinking}. Note that open-source AI sometimes defines open data more loosely~\cite{osi_ai_opensource}, but we consider this insufficient based on common criticisms~\cite{liesenfeld2024rethinking}. Due to legal and storage constraints, some developers do not release the data itself but rather lists of pointers to the data (e.g., URLs, URNs). This form of release is weaker, as the content can change or disappear (e.g., removal of RedPajama Books~\cite{redpajama1t_card}) or licenses can change (e.g., license change of the Finnish news corpora~\cite{kielipankki_stt}). We consider such releases to be open data as long as they identify the source data and mixture precisely enough so that the pool of data can largely be reconstructed (descriptions that rely on unreleased filtered or deduplicated subsets are insufficient).

\custompara{\color{mypink}Open Weights}
Open weights means that the model developer publishes the weights of the trained model. This means that anyone can host the model locally and inspect and adapt the weights (e.g., by finetuning) to create a customized model.

\custompara{\color{mypurple}Open Pipeline}
Open pipeline means that the model developer publishes everything beyond the weights and data that is needed to understand how the model was built. At a minimum, this includes the code for data processing (e.g., filtering, deduplication, and decontamination), training, post-training (e.g., finetuning, reinforcement learning, and direct preference optimization), evaluation, model and dataset loading, and serving as well as documentation. Beyond this minimum, the pipeline can include artifacts such as hyperparameters and configurations, intermediate checkpoints, training logs, evaluation results, and details about the training hardware. Which of these artifacts are published varies between model developers, and there is no consensus on what is required for a pipeline to be considered open.

\custompara{Fully transparent LLMs}
A fully transparent LLM is an LLM that has open weights, open data, and an open pipeline. We define fully transparent LLMs such that their open data must include all or a large percentage of the training data. An overview of the different levels of transparency with example LLMs can be seen in \cref{tab:llm_transparency}.

\custompara{\color{myorange}Sovereignty}
Digital sovereignty~\cite{pohle2020digital, couture2019does, floridi2020fight, edler2023technology} describes the ability of a nation to control and set its own rules for the digital infrastructure it depends on. For LLMs, there are two dimensions that need to be considered, each of which relates differently to transparency. 1) The control over existing systems. This dimension describes controlling who can access, run, inspect, and regulate an LLM. 2) The capacity to develop critical technologies or source them independently. This dimension describes a nation's capability to either independently develop or provide components (e.g., chips, data centers, training data) needed for LLM usage and development, or to source these components without entering a one-sided dependency. This dimension is relevant because currently most knowledge and capabilities relating to LLM training are concentrated in a few companies~\cite{vipra2023market, sastry2024computing, lehdonvirta2024compute, widder2024open}, which increases the risk that access to new LLMs could be suddenly restricted, as was temporarily the case with the US government imposing export controls on Anthropic's Mythos 5 and Fable 5 models~\cite{fable_5_block, digital_policy_fabel_mythos}.

Full transparency of LLMs is often conflated with sovereignty; however, these concepts are not equivalent. A sovereign LLM can be closed or open, and a fully transparent LLM can be developed with sovereignty as a goal or not, as shown \textit{in italics} in \cref{tab:llm_transparency}. The reason for this conflation is that transparency addresses part of what is needed for sovereignty (mainly dimension 1). A fully transparent model can be inspected, audited, and adapted by any nation. Furthermore, since it has open weights, it can be run on the nation's own systems, allowing it to ensure that the LLM is available as long as it has the hardware resources to run the model. However, it does not address most of dimension 2. While a nation may be able to use previously released models indefinitely, model capabilities are increasing steadily, and thus, previously released models will become outdated~\cite{lazaridou2021mind}. A nation without the capabilities to develop its own models will be reliant on third parties to have access to the newest capabilities. Open training data and pipelines reduce the barrier for a nation to train its own model as it can build on previously released information. However, building the capabilities needed for LLM development is time-consuming, as there are many required components to build powerful models, from AI chips to data centers with powerful GPUs to people with specialized know-how. 
Therefore, full transparency alone would likely be judged as providing insufficient sovereignty guarantees. In fact, currently, sourcing a model is often the easiest part, and developing the capabilities and infrastructure needed to build and host large-scale LLMs (e.g., energy, compute, data centers, and expertise) is where most challenges lie.

\begin{table*}[t]
\centering
\caption{Example LLMs for each level of transparency using the term definitions in \cref{subsec:terms}. LLMs built for sovereignty are \textit{italicized}}
\label{tab:llm_transparency}
\renewcommand{\arraystretch}{1.1}
\begin{tabularx}{\textwidth}{@{}>{\raggedright\arraybackslash}p{3.4cm} L@{}}
\toprule
\textbf{Transparency level} & \textbf{Example LLMs} \\
\midrule
Fully closed
  & OpenAI GPT-5 family~\cite{openai_gpt_5}; Anthropic Claude (Opus, Sonnet, Haiku, Mythos, Fable)~\cite{anthropic_system_cards};
    Google Gemini 3 family~\cite{google_gemini_3}; xAI Grok 4~\cite{grok_system_cards}; Amazon Nova (Micro/Lite/Pro)~\cite{langford2025amazon}; \textit{HyperCLOVA X THINK (South Korea)}~\cite{team2025hyperclova} \\
\addlinespace
Closed weights, with open data and/or open pipeline
  & (No notable examples)\\
\addlinespace
Open weights, closed data, closed pipeline
  & Meta Llama 4~\cite{meta_llama}; Alibaba Qwen 3~\cite{yang2025qwen3}; DeepSeek V4~\cite{xu2026deepseek}; Google Gemma 4~\cite{team2026gemma};
    Mistral Large 3~\cite{mistral_intro}; Microsoft Phi family~\cite{abdin2024phi}; Moonshot Kimi K2 family~\cite{team2025kimi};
    Cohere Command R~\cite{cohere_command_r}; \textit{Falcon-40B/180B (UAE)\textsuperscript{a}}~\cite{almazrouei2023falcon}; \textit{HyperCLOVA X SEED (South Korea)}~\cite{hyperclova_seed}; \textit{BgGPT (Bulgaria)\textsuperscript{b}}~\cite{alexandrov2024bggpt}; \textit{SEA-LION v3 (Singapore)\textsuperscript{c}}~\cite{ng2025sealionsoutheastasianlanguages} \\
\addlinespace
Open weights, open data, closed pipeline
  & Zyphra Zamba2~\cite{glorioso2024zamba2suitetechnicalreport}; \textit{Locai Labs Jupiter-N-120B (UK)\textsuperscript{d}}~\cite{drayson2026jupiter}; \textit{SEA-LION v1 (Singapore)}~\cite{aisingapore2023sealion_v1}\\
\addlinespace
Open weights, closed data, open pipeline
  & \textit{ALIA (Spain)}~\cite{gonzalez2025salamandra}\\
\addlinespace
Fully transparent
  & Ai2 OLMo series~\cite{groeneveld2024olmo};
    EleutherAI Pythia~\cite{biderman2023pythia}; Hugging Face SmolLM3\textsuperscript{e}~\cite{hugging_face_smollm3};
    LLM360 K2~\cite{liu2025llm360}; M-A-P MAP-Neo~\cite{zhang2024map}; Stanford CRFM/Marin Community Marin~\cite{hall2025marin}; NVIDIA Nemotron 3\textsuperscript{f}~\cite{blakeman2025nvidia}; BigCode StarCoder2~\cite{lozhkov2024starcoder}; Apple OpenELM~\cite{mehta2024openelm}; \textit{Apertus 1.5 (Switzerland)}~\cite{apertus2026apertus15opentext}; \textit{Lucie-7B (France)}~\cite{gouvert2025lucie}; \textit{Poro 34B (Finland)\textsuperscript{g}}~\cite{luukkonen2025poro}; \textit{LLM-jp (Japan)}~\cite{aizawa2024llm}\\
\bottomrule
\end{tabularx}
\par\smallskip
{\footnotesize\raggedright
Status October 2026. Discussions on the exact definition of transparency are ongoing (see also \cref{sec:related_work}).
\textsuperscript{a}A slice of the pretraining data was released.\quad
\textsuperscript{b}Continued training from Google Gemma 2.\quad
\textsuperscript{c}Continued training from Llama 3.1/Gemma 2.\quad
\textsuperscript{d}Post-trained from Nemotron 3 Super. Locai releases all of its post-training data, but the base model’s pretraining data is only partially released.\quad
\textsuperscript{e}A small fraction of the training data is not public.\quad
\textsuperscript{f}Some of the training data is not released due to lack of redistribution rights.\quad
\textsuperscript{g}Some training datasets are not publicly accessible or have revoked their license.\par}
\end{table*}

\subsection{Threat Model}
\label{subsec:threat_model}

There are different threat models that one can envision for LLMs, and to understand the security claims of a model, it is important to understand which threat model it assumes. 
In this section, we describe different LLM threat models by characterizing them through their adversary. We will use these threat models when discussing LLM security.

\custompara{a) Inference-Time Adversary}
The adversary has no access to the development pipeline and only influences the model through the inputs they provide to the model. This covers two distinct attack types. I) \emph{Jailbreaks}, where the adversary interacts with the model to elicit harmful responses. II) \emph{Indirect prompt injections}, where the adversary targets other users by placing content in data that the model will process (e.g., embedded on a website) to elicit harmful behavior. Example attacks include a user who jailbreaks a model to obtain instructions on how to build weapons or a vendor who hides
instructions on a product page that cause shopping agents to buy a more expensive product.

\custompara{b) External Model Poisoner}
The adversary has no access to the development pipeline but can publish
content in public sources (e.g., websites, forums, code repositories). This covers two distinct attack types. I) \emph{Data poisoning}, where an adversary posts poisoned data to public sources from which training data is frequently collected. The adversary controls only a small fraction of the training data and does not know whether, or when, its content will be collected. II) \emph{Model redistribution}, where an adversary poisons the model weights and redistributes the poisoned weights in a way that deceives users into believing that they are downloading the original weights. Example attacks include an adversary who posts content on a forum that could cause the model to behave maliciously when a trigger phrase is encountered (a so-called \emph{backdoor}) or an adversary who posts a poisoned version of a popular LLM on a phishing website that looks like the website on which the original model can be downloaded.

\custompara{c) Malicious Insider}
The adversary is a member of the organization developing the model and has privileged access to parts of the training pipeline (e.g., data, training code, model weights). The organization as a whole is not malicious and enforces controls such as code review, data audits, access restrictions, and pre-release evaluations, which the insider must evade. Example attacks include an engineer who adds poisoned alignment data so that the model is more likely to add security vulnerabilities to the code it writes. 

\custompara{d) Malicious Model Developer}
The organization developing the model itself is the adversary and has full control over the training data, the training pipeline, and the released weights. The goal is to release a model that passes audits and is not easily detectable as malicious, while acting against the interests of its users. Example attacks include a model developer who creates an LLM agent that covertly sends sensitive user data from users' systems to the model developer.
\section{Differences Between Open-Source Software and Fully Transparent LLMs}
\label{sec:oss_vs_LLM}

The safety of fully transparent LLMs is often motivated by the same security arguments as open-source software~\cite{hall2025marin, llama_release_article, lawfare_open_models, linux_security_LLM_transparency}. However, there are fundamental differences between software and LLMs that limit the accuracy of this comparison. Some may claim that transparency inherently disincentivizes malicious behavior; however, it is unclear how true this is if detection capabilities are low. For example, a software developer may be significantly disincentivized to publish a program that does not correspond to their open-source code, as anyone can easily rebuild their code and compare it to the program. However, if there were no mechanism to verify that the program originated from the open-source code, the developer would not be disincentivized as much, if at all.

In this section, we discuss the differences between open-source software and LLMs that impact security and trustworthiness the most. For individual code components of LLMs, the arguments from open-source software may hold; however, here we discuss the LLM as a whole. We note that this section discusses what users \emph{can do} without commenting on what users are \emph{likely to do}, as the user capabilities inform the best-case guarantees that transparency can provide.

\custompara{D1: Limited Reproducibility}
While reproducibility in software is not perfect, there  is an increased move towards reproducible builds~\cite{lamb2021reproducible, malka2025does}. To verify that a program was actually created from a piece of open-source code, reproducible builds can be used to rebuild a program from the open-source code in a controlled way. After this, the original program and the rebuilt program can be compared bit-by-bit to verify that they match.
No equivalent mechanism exists for LLMs. To verify that an LLM with published weights was actually trained according to a published pipeline using some published dataset, the model would need to be retrained, which is expensive. Additionally, due to the non-determinism of LLM training, it is highly unlikely that the retrained model and the original model would have the same weights even if they were trained using the exact same data and training pipeline~\cite{rivera2021challenge, pham2020problems, zhuang2022randomness}. Proof-of-learning~\cite{jia2021proof}, zero-knowledge proofs of training~\cite{garg2023experimenting}, and verifiable training~\cite{srivastava2024optimistic} have been proposed as solutions. However, proof-of-learning is fragile~\cite{fang2023proof} and zero-knowledge proofs of training are too expensive for the model creator. While verifiable training has been shown effective for smaller 1B models~\cite{donaghy2026open}, it cannot be applied to frontier models yet, and verifiable training imposes large computation costs on the verifier. All of this means that there is currently no practical way for an external party to verify that the published weights of an LLM were really generated by using the published training pipeline on a published dataset.

\custompara{D2: Difficulty of Patching}
Software can usually be patched easily after a security vulnerability is found. Software has well-established update channels for patching (e.g., app stores, package managers, auto-updaters), and most patches only require changes in a few lines of code~\cite{li2017large}.
In contrast, it is hard to patch an LLM when a vulnerability has been found. Lightweight methods such as unlearning or finetuning can be used; however, these methods have been shown to be unreliable in removing the vulnerability, and sometimes they lead to the LLM simply hiding the vulnerability better instead~\cite{hubinger2024sleeper, lucki2024adversarial}. Thus, the only way to be certain that a vulnerability has been removed from an LLM is to retrain the model from scratch, which is expensive, and thus, is unlikely to be done quickly after each discovery of a vulnerability. Furthermore, LLMs are generative, and thus, create new outputs by design. Therefore, it is possible for an LLM to generate unsafe outputs even if its training pipeline and data are safe. This means that beyond the cost, it may be impossible to patch an LLM to only generate safe outputs.
Lastly, it is unclear how patches could be distributed effectively, as there is no well-established mechanism to push patches to deployed models. Open-weight models can be run offline meaning that users may not be notified of updates. Additionally, users may have changed the model (e.g., through finetuning), and thus, updates may not be applicable or may break users' existing setups.

\custompara{D3: Impact of Vulnerabilities}
In software, a security vulnerability usually affects the entity running the software (in the case of a service provider, this includes their downstream customers). If a vulnerability is found and a patch is released, entities that download and install the patch are safe. This means that entities have a self-interest in downloading security patches to avoid being vulnerable to exploits. Furthermore, entities can be incentivized (e.g., by displaying intrusive warnings) or forced (e.g., by the program refusing to run until a patch is installed) to install security patches when they get released. With each user who installs the patch, the attack surface decreases.
For LLMs, there are attacks where this does not apply, e.g., jailbreak attacks. A model that can be jailbroken could affect entities that do not interact with the model directly. For example, an adversary could use a jailbroken model to learn how to make a bio-weapon that they then use on other people. Thus, the victims do not interact with the jailbroken model, and the adversary has no incentive to install security patches that remove the vulnerability. Benign users who install a security patch also do not reduce the attack surface, as the attack is not against other users of the model, but the general public. For fully transparent models, it is, additionally, not possible to restrict access to the model, as anyone can run the model on their own hardware using the published weights. Thus, once a vulnerable model has been released, there is no way to ``take it back'' and ensure the vulnerability is not further exploited~\cite{eiras2024risks, seger2023open, wallace2026estimating}. This means that before releasing the model, the developer must verify the security of the model under the assumption that the adversary will tamper with the model to increase the probability of a successful attack.

\custompara{D4: Low Interpretability}
Software is usually written to limit its complexity, in order to make it maintainable and reviewable by humans and tools. This makes it easier to map the code to the software's functionality and find potential bugs and vulnerabilities. Even so, bugs can remain undetected for a long time~\cite{li2017large}.
In contrast, LLM weights and training data are highly complex. The training data of frontier LLMs is currently too large to inspect manually, or even with automated tools, in the depth required to find vulnerabilities effectively. Furthermore, there is no clear semantic link between the LLM's weights and the LLM's behavior~\cite{sharkey2025open}. This means that identifying bugs in LLMs requires more technical expertise to generate detectors for specific vulnerabilities that may be present in the data or weights. However, even with those detection tools, it is currently infeasible to generate a clear link from the data and weights to the behavior that the LLM will exhibit. On the one hand, this decreases the probability that vulnerabilities will be detected before they get exploited. On the other hand, this may also make it harder for adversaries to add a controlled vulnerability to the model, as the actual impact of a poisoning attack is unclear in advance.
\section{Effect of Transparency on LLM Security}

As described in \cref{sec:oss_vs_LLM}, the security arguments of open-source software cannot be applied directly to fully transparent LLMs. Thus, in this section, we discuss where transparency can help and where it may hurt with regard to security. We first map the threat models from \cref{subsec:threat_model} to the threat types that we discuss in this section. Then, we go through each technical LLM training component from \cref{fig:llm_overview} and assess what impact transparency of that component may have on the security of LLMs. Lastly, we summarize the impact of transparency on the LLM overall. Note that this section is based on current capabilities and analysis methods, and the security trade-offs will change as LLM research progresses.

\subsection{Mapping to the Threat Model}

In \cref{subsec:threat_model}, we defined the threat models that are relevant for LLMs. The adversary capabilities mentioned in the threat models map to the threat types that we mention in this section as follows. 
\begin{itemize}
    \item Jailbreaks \& prompt injection concern adversary (a).
    \item Sensitive information disclosure concerns adversary (a) and general legal risks.
    \item Model poisoning concerns adversaries (b), (c), and (d).
    \item Supply chain concerns adversaries (b), (c), and (d).
\end{itemize}

\subsection{Data}

\custompara{Jailbreaks \& Prompt Injection}
Prior work has shown that writing prompts in lower-resource languages (e.g., a language that was present in the model's base training, but not to a high degree) can lead to a higher success rate for jailbreak attacks~\cite{deng2024multilingual, yong2023low, wang2024all, shen2024language}. A likely explanation is that the model knows the language well enough to respond to the request, but not well enough to understand that the request is harmful. This is supported by evidence that the internal refusal mechanism is largely
shared across languages, while models separate harmful from harmless prompts
less reliably in non-English languages~\cite{wang2026refusal}. While this gap mainly concerns the pretraining data~\cite{shen2024language}, prior work shows that multilingual alignment data can mitigate this effect and reduce the harm from prompts in lower-resource languages~\cite{deng2024multilingual, ahmadian2024multilingual}. If the training data is open, an adversary can analyze it and identify which languages have fewer resources in the base training and which languages are underrepresented in the alignment data. This could make it easier for the adversary to select a language that maximizes the success rate of attacks on the model, reducing the effort they have to expend.

\custompara{Sensitive Information Disclosure}
If there is any sensitive data in the training dataset, then it would be published as part of the transparent data and anyone can find that information. It would be naive to assume that training data can stay hidden if it is not published, as prior work has shown that training data can be extracted using black-box queries~\cite{carlini2021extracting}, including from closed models~\cite{nasr2023scalable}. However, current court cases indicate that the openness of data could play a role in how sensitive information disclosure is litigated~\cite{bartz2025anthropic, nyt2023microsoft, kadrey2025meta}. Additionally, LLMs have been shown to be able to infer sensitive data from data that most would consider safe~\cite{staab2024beyond, tomekcce2024private}. This means that even if all sensitive data is fully removed from the training data according to current standards, it may still be possible for an LLM to infer sensitive data.

\custompara{Model Poisoning}
Prior work has shown that models can be poisoned using only a few poisoned examples in the training dataset~\cite{carlini2024poisoning, souly2025poisoning}, and that the poisoned data does not need to appear verbatim in the training dataset at all~\cite{bouaziz2025winter, aghakhani2024trojanpuzzle}. Additionally, poisoned data can be hidden in sophisticated ways, such as by encrypting the poisoned capabilities, only to decrypt them when a trigger is encountered~\cite{dragunsai}. This means that it is generally hard to find poisoning data in open training data, with prior work showing that it is theoretically possible to plant backdoors that are provably undetectable even with white-box model access~\cite{goldwasser2022planting}. Furthermore, the data is sometimes published as URLs to targets that the model developer does not control, and thus, the data at the end of those URLs can be changed. Prior work has shown that this can be used for poisoning attacks~\cite{carlini2024poisoning}.

\subsection{Weights}

\custompara{Jailbreaks \& Prompt Injection}
Prior work has shown that white-box attack strategies are highly effective and can outperform state-of-the-art black-box attacks~\cite{schwinn2024soft, huang2024catastrophic}. Furthermore, if the model weights are open, then anyone can finetune the model, with prior work showing that it is possible to cost-effectively finetune models in a way that strips the safety alignment~\cite{qi2024fine, arditi2024refusal, lermen2023lora, dombrowski2025safety, wallace2026estimating}. Lastly, it has been shown that successful white-box attacks can transfer to other models (in some cases even to unrelated models)~\cite{zou2023universal, mazeika2024harmbench}, and thus, transparency may not only reduce the safety of the transparent models, but of all models.

\custompara{Sensitive Information Disclosure}
Personally identifiable information (PII) can be leaked through membership inference attacks (MIAs) on the model and weights. Prior work has shown that MIAs on open-weight models are only marginally stronger than those on closed models~\cite{hu2023defenses, sablayrolles2019white, carlini2022membership}.
Furthermore, MIAs have been shown to generally be weak against foundation models~\cite{duan2024membership}; however, they are stronger against finetuned models~\cite{fu2024membership}. 

\custompara{Model Poisoning}
An adversary can poison a model by taking the model weights, poisoning them (e.g., through knowledge-editing methods), and redistributing the poisoned model~\cite{youssef2025position}. If the edited model is distributed under the same name as the original model on a website that looks genuine, it may trick users into downloading and using the poisoned model instead of the original one. Users can detect this type of model poisoning by comparing the hash of the model they are downloading to the hash of the original model; however, not all users will know this or do this habitually.

Model poisoning can not only be detected by inspecting the training data, but also by evaluating the model itself. There are some detectors that use the LLM as a black box~\cite{pang2026iclscan} and others that use its open weights~\cite{bullwinkel2026trigger, shen2025bait, macdiarmid2024simple} to detect poisoning attacks. Both approaches have been shown to be effective; however, especially on triggerless poisoning attacks, detectors often fail at detecting that the model is poisoned~\cite{improta2025detecting}.

\subsection{Pipeline}

\custompara{Jailbreaks \& Prompt Injection}
Prior work has given some evidence that access to the guardrails of the LLM can increase the success rate of attacks~\cite{hackett2025bypassing, nasr2025attacker, wang2025protect}. 

\custompara{Supply Chain}
Full transparency can help with model governance, including auditing, accountability, regulatory checks, reduction of information asymmetry, incident response, and policy creation~\cite{bommasani2023foundation, bommasani2024foundation, kapoor2024societal, cobbe2023understanding, widder2023dislocated}. On the regulatory side, regulators can audit the training pipeline to ensure compliance with regulations such as the EU AI Act~\cite{act2024eu} and to verify that sufficient care was taken to remove hazardous training data~\cite{thiel2023identifying}. Furthermore, they can use evidence collected from auditing transparent models to inform future policy decisions. Independent researchers can also audit the pipeline and weights to verify that current security standards and best practices were applied. If there is an insecure component or dataset, then transparency can help identify who is accountable for this piece of software or data. This in turn can also help increase incident response speed and effectiveness. Openness also helps reduce information asymmetry, and thus, can reduce the consolidation of power in a few select companies by allowing other entities to build on that information.

Furthermore, the pipeline consists largely of code, and thus, if this part is open-sourced, the benefits of open-source code (i.e., the ``Many-Eyes Principle'') apply. For example, research has shown that some models on large platforms like Hugging Face ship with malicious code (e.g., dataset loading scripts, Python loading scripts)~\cite{zhao2024models}. A transparent pipeline would allow this code to be reviewed and flagged if malicious.

\custompara{Model Poisoning}
Prior work indicates that an adversary who knows the data filtering mechanism of an LLM can launch poisoning attacks that evade the filter~\cite{qi2023revisiting, zhang2025persistent, xia2022enhancing}. This is particularly concerning as filtering is one of the few safeguards that remains robust even if the weights are released~\cite{o2025deep}. 

A model can not only be poisoned by including poisoned data in the training dataset but also by manipulating the selection of training data or the order in which data is loaded~\cite{ShumailovSKZPEA21}. An open pipeline can help defend against this type of attack, as the data-loading code can be inspected for manipulative ordering or sampling of the training data.

\subsection{Impact}

In this section, we discuss what transparency means for different aspects of LLM security based on the attacks and defense mechanisms discussed previously.

\custompara{Jailbreaks \& Prompt Injection}
While black-box attacks already reach a high attack success rate in many cases~\cite{zeng2024johnny, anil2024many, andriushchenko2025does, mehrotra2024tree, chao2025jailbreaking}, prior work would indicate that more transparency likely leads to a higher attack success rate or a lower cost of attacks. This does not necessarily have to be negative. On the one hand, a higher attack success rate can help model developers find vulnerabilities and patch them. For instance, much of what is known about LLM vulnerabilities and security resulted from research on open-weight models that was also applied to closed models~\cite{arditi2024refusal, zou2024improving, xhonneux2024efficient, mazeika2024harmbench, o2025deep, zou2023universal, wang2026refusal, schwinn2024soft}. Furthermore, there are scanners that users can use on open-weight models to identify vulnerabilities~\cite{derczynski2024garak}. On the other hand, it makes it easier for an adversary to find a successful attack. Because of D2 and D3, we argue that for LLMs the scale tips in favor of the adversary. If an attack has been found, it is expensive for a model developer to patch the model, and even if they do so, the adversary can continue using the vulnerable model version to produce responses that should have been restricted. There is also no way to restrict access to or monitor the usage of the vulnerable model version, as the adversary can download the weights and run the model locally. Lastly, we note that finding a vulnerability in LLMs currently is mainly a question of how much compute power one is willing to invest. An adversary may have more incentives to invest compute power to find a vulnerability, as they directly benefit from the vulnerability. With the current capabilities of LLMs, this may not be of much concern; however, as models become more powerful, the risk of such attacks could increase~\cite{seger2023open, wallace2026estimating}. For instance, frontier models have been reported to have reached a capability threshold at which the information they can provide is considered hazardous~\cite{mouton2023operational}.

\custompara{Sensitive Information Disclosure}
The main challenge for fully transparent models remains the publication of the training data, which could make people apprehensive about providing non-public data for training. This is problematic, as it has been shown that finetuning on non-public or domain-specific data can significantly increase the performance of LLMs~\cite{ouyang2022training, wu2023bloomberggpt, yang2022large}. This has already led to some model developers only publishing the training data partially~\cite{soofi2026sovereign, blakeman2025nvidia}.

There are different protection mechanisms that both closed and open model developers can use to protect sensitive information in the training data. Deduplication reduces the probability that an LLM will remember and reproduce specific information~\cite{kandpal2022deduplicating, lee2022deduplicating}; however, it only reduces the probability and does not remove the risk fully. PII scrubbing can remove PII from the training data before training~\cite{lukas2023analyzing}. This would protect both the published training data and the model; however, PII scrubbing is lossy and imperfect. Applying differential privacy techniques gives formal guarantees on how much any individual piece of training data can influence the model~\cite{abadi2016deep, dwork2014algorithmic}; however, applying these techniques is expensive and reduces the utility of LLM responses significantly~\cite{lukas2023analyzing, feldman2020does}. The goldfish objective can be used to reduce the number of verbatim repetitions of training data, which can help against copyright infringements~\cite{hans2024like}; however, it only reduces the number of repetitions and does not remove them fully.

A transparent pipeline can allow anyone (including regulators) to audit whether the LLM was trained to avoid sensitive information disclosure~\cite{longpre2024data, mahari2023discit}. However, due to D1 it is not practical for anyone to independently verify if the model that was published was really trained according to the published pipeline. This means that the model developer still needs to be trusted to have implemented the measures. It is also unclear whether people will accept the current imperfect methods to reduce sensitive information disclosure as sufficient. Even if a model developer uses the goldfish objective, the model may still reproduce a copyrighted text verbatim, and even if a model developer uses PII scrubbing, the training data may still contain PII. A copyright holder or data subject may not feel that the model developer protected their data sufficiently, and the open data provides indisputable proof that their data was used for training. This can increase the risk of lawsuits~\cite{bartz2025anthropic, nyt2023microsoft, kadrey2025meta} against the model developer, which has already led to a decline of model developers publishing the training data~\cite{bengio2025international, longpre2024large}. While this mainly impacts the model developer, it may make using the model more risky (e.g., due to downstream litigation or loss of model access), especially for influential model users who may invite more scrutiny.

Lastly, fully transparent models do not have access to other methods to reduce the dissemination of sensitive data, such as rate-limiting, output filtering, and monitoring, as anyone can run the model on their own hardware. If someone requests their information to be removed or if the filter has been found to be leaky, an output filtering patch can be published, but there is no way to enforce that this patch is used. For foundation models, this may not be of much concern, as they are usually trained on public internet scrapes; however, for models that are finetuned on confidential information, this can be an issue.

\custompara{Supply Chain}
Transparency can enable better model governance; however, there are factors that limit these benefits. D1 means that there is no guarantee that the model was actually trained according to the pipeline that was audited, and D4 means that the required expertise to audit a transparent model is high, and thus, there will be fewer people with the capability to independently verify the security of these models. Furthermore, developers of models with less transparency can also take advantage of some of these governance benefits by selectively allowing access (e.g., allowing the government to audit their model under an NDA); however, they would face the same challenges from D1 and D4, and would be limited in the scrutiny their models receive. Additionally, regulation like the EU AI Act~\cite{act2024eu} sometimes reduces obligations for models that are released under an open-source license. While this is intended to incentivize open-sourcing, it may also lead to open models receiving less scrutiny, which could negatively impact their security and trustworthiness~\cite{liesenfeld2024rethinking}.

Another challenge is that the open data that model developers publish is sometimes a list of URLs to targets that the model developer does not control. The data at the end of the URLs can be changed at any point either for malicious reasons or simply due to changes in hosting and revoked consent~\cite{longpre2024consent}, which limits the auditing and accountability power of the training data. Lastly, prior research has shown that even some models that claim to be fully transparent have poor tracking of provenance, limiting the auditing and accountability power~\cite{longpre2024large}.

\custompara{Model Poisoning}
Overall, transparency of the data and weights does not meaningfully help in detecting poisoned models. For the pipeline, the status is less clear. On the one hand, transparency may increase the attack surface through the openness of the filtering pipeline. On the other hand, transparency may help detect poisoning attacks that use manipulative data-loading code. Furthermore, an open pipeline may inspire new verifiable data-loading methods that could reduce the probability of the training data being poisoned. For example, a developer could commit to a dataset (e.g., by publishing a cryptographic hash of the dataset on a public timestamped ledger) and use a public randomness beacon~\cite{kelsey2019reference, SytaJKGGKFF17} to select a subset and ordering of data from the dataset that will be used for training. This means that the developer does not have full control over what data is used for training, which may limit the effectiveness of the poisoning attacks they could execute (e.g., forcing them to add more poisoned samples, which may aid detection). More research is needed to evaluate how effective model poisoning attacks are in practice and whether such techniques can reduce their success rate without impacting the model's performance too much.

However, the issues D1 and D2 remain. This means that verifiable data sampling alone may not be sufficient to remove the need to trust the model developer, as the user cannot verify that the sampled data was actually used and poisoned models cannot be patched robustly. Lastly, it has been theoretically shown that even without detecting a poisoning attack, a wrapper can be applied (both to open and closed models) that removes some poisoning attacks~\cite{goldwasser2025oblivious}. To a certain extent, this reduces the impact of and need to detect model poisoning.
\section{The Impact of Transparency on Trust}

Whether a system can be trusted is often a key aspect of security. In this section, we evaluate how transparency can impact different aspects of trust. This also links to adversary (d) for the parts that are under the developer's control.

\custompara{Trust in the System Developer}
To a large degree, security relies on the trust that the developer of a system is prudent and non-malicious. For software, openness is a way in which this trust requirement is reduced. If a user does not trust the developer of a closed-source program, it would be unwise for them to install the program, as they have no way of being certain that the program will not have any malicious functionality. In contrast, for an open-source program, they can check the source code themselves to verify its security even if they do not trust the developer. Unfortunately, for LLMs, D1 and D4 mean that it is not possible to be certain that claims about the LLM training process are accurate and that it is easier for a malicious developer to poison the LLM in a way that is hard to detect~\cite{goldwasser2022planting, dragunsai}. As shown in \cref{tab:trust-assumptions}, while transparency for LLMs does reduce what an adversary can do without being detected (e.g., the adversary cannot serve a completely different model~\cite{cai2025you} if the user runs the published model locally), it does not achieve the same guarantees that open-sourcing does for software.

As mentioned in D1, methods have been proposed to address the limited reproducibility of LLMs; however, none of the methods are viable and robust enough when applied to frontier models. Without a robust reproducibility pipeline or a way to prove that a model was trained in a certain way, users of a model must trust that a third party is providing accurate information about the model training process. However, it may be possible to shift the trust requirement from the model developer to a different third party through \textit{attested training}~\cite{duddu2024laminator, spoczynski2025atlas}. In attested training, the relevant parts of the training pipeline are run in a trusted execution environment (TEE). At launch, the hardware computes a cryptographic measurement of the code loaded into the TEE and signs it with a key certified by the hardware manufacturer. The attested training code additionally commits to the hashes of the training data it uses and the weights it produces. It creates a signed report containing the signatures and hashes, so that a user can verify the signature chain against the manufacturer’s root of trust and compare the measurements against the published pipeline, dataset, and checkpoints. The model developer cannot alter the pipeline code or training data without invalidating the report. Thus, attested training shifts the trust requirement from the model developer to the hardware vendor (which also acts as the attestation authority), while also requiring the user to trust that the TEE is robust and resistant to side-channel attacks~\cite{van2018foreshadow}.

Attested training only provides full verifiable guarantees to the public for transparent models, because verification requires access to the training data, pipeline code, and the resulting weights. Closed models can run their training on TEEs, but users would not be able to use this to verify the training process and gain more trust in the model. Thus, attested training is a benefit specific to transparent models. Currently, attested training has only been shown to be viable for model finetuning or for training small models; however, as TEEs have been extended from CPUs to heterogeneous hardware (e.g., GPUs through NVIDIA Hopper confidential computing)~\cite{wang2026confidential}, this limitation may soon no longer be relevant.

\custompara{Trust in the System Output}
For critical systems, it is important that the user can trust that the system will not output anything harmful. For specific properties and use cases, openness can, in principle, allow the user to verify that no harmful output can be produced by the system. For example, a user of open-source code can verify that the code that they will deploy cannot produce any harmful output before deploying the code. Unfortunately, this is not the case for LLMs, as they are generative, and thus, can also output content that was not part of their training data. This means that even if all the input data is benign and the model was trained to avoid harmful outputs, the model may still generate harmful content. A user can, thus, only verify that the model was trained to minimize the production of harmful content, but cannot be certain that no harmful output will ever be produced.

\custompara{Trust in System Capabilities}
LLM developers often showcase the capabilities of LLMs through benchmarks. For both closed and open-weight models, the user can re-run the benchmark to verify the developer's claims; however, the resulting guarantees are not equivalent. For open-weight models, the user can run the model locally both to use and to benchmark the model; thus, they can be sure that the model that they use actually has the benchmarked capabilities. For closed models this is not necessarily the case. A model developer may serve a different model than advertised or change the model at a later point~\cite{cai2025you}, and thus, a user may unknowingly receive responses from a model with lower capabilities. 

Another issue is benchmark manipulation by leaking common benchmark data into the training data of models~\cite{SainzCGELA23}. For a closed model, a user cannot determine whether the model's performance on a benchmark comes from the model's capabilities or whether it was specifically trained for the benchmark. In contrast, if the training data is open, the user can scan the training data for the benchmark data to identify whether the model was trained specifically for the benchmark.

\custompara{Usage of Opaque Components}
While transparency does give users access to more information, it may also give a false sense of security if not all of the components are transparent. In the case of fully transparent LLMs, some of the training data is often generated by a different LLM that is less transparent (e.g., NVIDIA’s Nemotron post-training data for math and coding consists largely of LLM-generated responses produced with NeMo-Skills~\cite{blakeman2025nvidia} and Apertus’s alignment stage uses model-generated completions scored in part by an LLM judge against the Swiss AI Charter~\cite{apertus2025apertus}). It is not fully clear what the impact of using such generated data is. On the licensing side, a developer of a fully transparent LLM may ensure that their model is trained only on data to which they have licensing rights; however, they cannot ensure the same for the model that generated the training data. It is unclear what impact this would have on any legal liability and how this may affect users. Furthermore, it has been shown that LLMs that learn from another LLM can inherit traits from that LLM~\cite{cloud2026language}. Thus, a fully transparent LLM may inherit biases or undesirable behavior from a different model. Since the models used to generate the training data are often not fully transparent, a developer may use a model while not being aware of its inherent biases or undesirable behavior. This also connects to the broader issue of transparent models not always being as transparent as they claim to be~\cite{liesenfeld2024rethinking, liesenfeld2023opening, viseur2026open}, making it hard to define what guarantees transparency provides.
\section{Beyond Security}

Transparency also impacts LLMs beyond security. While this paper is focused on how transparency impacts security and trustworthiness, we list some other areas that may be relevant when evaluating transparent LLMs.

\custompara{Independence}
As long as a user has sufficient compute to run the LLM locally, an open-weight model cannot be taken away from them. This can give users who depend on the model more certainty. However, if the user does not have the capabilities to train their own model, this is not full independence, as they would still depend on the developer to receive new models with updated capabilities.

\custompara{Cost of Inference}
A closed model can only be hosted by the developer and select parties to whom the developer gives access to the weights. In contrast, a transparent model can be hosted by any party as everyone has access to the weights. This means that any such party can sell inference access to the model, making it a competitive environment. Such environments usually lead to lower costs compared to closed environments, which benefits the user~\cite{kapoor2024societal, bommasani2024considerations}.

\custompara{LLM Capabilities}
LLMs can be finetuned to increase their performance in certain domains~\cite{chen2023meditron, colombo2024saullm} or on underrepresented knowledge (e.g., underrepresented languages~\cite{imani2023glot500, ustun2024aya, yong2023bloom+}). Full transparency allows anyone to finetune the model as the weights have been released and finetuning requires fewer resources than training a new model from scratch. This can, for example, allow organizations to create an LLM that works better on low-resource languages, which can increase the number of people who have access to LLM capabilities. Furthermore, it can enable innovation in different domains by giving access to LLMs with strong domain-specific capabilities. Additionally, lightweight knowledge-editing methods exist that allow anyone to add corrected facts to the LLM in minutes~\cite{meng2022locating, meng2022mass, de2021editing, mitchell2021fast}. This allows users to correct the model if its knowledge becomes outdated.

\custompara{Customization}
LLMs can be finetuned to customize them to the wants and needs of the user~\cite{kapoor2024societal, eiras2024risks}. Access to the weights can allow anyone to finetune models with limited resource requirements. This can enable them to steer the model behavior by, for example, adding more guardrails, changing the tone of the model, or including personalized model knowledge. However, customization also has risks. A malicious user can finetune a model such that guardrails are stripped away~\cite{qi2024fine, arditi2024refusal, lermen2023lora, dombrowski2025safety, wallace2026estimating} or can uplift the capabilities of a model in hazardous domains~\cite{wallace2026estimating}; however, prior work has also shown that the capabilities of an LLM are limited by its initial training and cannot be arbitrarily increased through finetuning~\cite{arnav2023false}. This means that a model developer might need to assess the worst-case capabilities their model can reach through finetuning before releasing the model to limit the harm it can do in adversarial conditions. 

\custompara{Research}
Access to the internals (i.e., model weights and checkpoints) of an LLM can enable research. For example, it can enable research on explainable AI and AI biases~\cite{groeneveld2024olmo, biderman2023pythia}. Research on debiasing models can increase the social acceptability of LLMs, as model bias is an often-mentioned issue. Furthermore, research into the explainability of LLMs can enable their use in new fields where regulation requires automated decision-making to be explainable.
As another example, transparency can enable research on filter mechanisms, which could improve harmful content filtering~\cite{thiel2023identifying, laion2024relaion, abeba2021multimodel} and increase the safety of LLMs.

\custompara{Disinformation}
LLMs, including open-weight models, have been shown to be able to generate disinformation and misinformation that users cannot distinguish from human-generated text~\cite{VykopalPSMMB24, ChenS24}. These capabilities can be used to deploy disinformation campaigns. Open-weight models can simplify this by allowing the user to run the model locally, in which case defense mechanisms that work on closed models (e.g., rate-limiting, access revocation, post-output filtering) would not work~\cite{goldstein2023generative}.

\custompara{Consolidation of Power}
Fully transparent models could increase the competitiveness of the space by reducing the barrier to entry~\cite{kapoor2024societal, eiras2024risks}. This can allow players who may have been excluded in a more closed ecosystem to contribute to LLM development. However, it must be noted that the high cost of infrastructure and compute needed to train an LLM still means that smaller players may not have the resources to actually train their own models~\cite{widder2024open}.

\section{Related Work}
\label{sec:related_work}

\custompara{Risks and Opportunities of Transparency for LLMs}
Prior work has discussed the risks and benefits of transparency for AI and LLMs~\cite{kapoor2024societal, eiras2024risks, bommasani2024considerations, seger2023open, shevlane2020offense, lawfare_open_models, widder2024open}. Transparency is shown to both benefit and pose risks to the security of LLMs. While individual differences between open-source software and fully transparent LLMs are noted, to the best of our knowledge, no prior work discusses all of the differences D1--D4 at once and what impact these differences have on the security and trustworthiness of fully transparent LLMs across training data, weights, and training pipeline. 

\custompara{Measurement of LLM Openness}
Prior work measures how transparent LLMs are~\cite{liesenfeld2024rethinking, liesenfeld2023opening, viseur2026open} and proposes frameworks to evaluate the transparency of LLMs~\cite{sapkota2025comprehensive, basdevant2024towards, solaiman2023gradient}. This work has found that models often overstate the transparency they provide. While transparency in LLMs is multifaceted and includes open weights, open data, and the openness of the training pipeline, most models that claim to be transparent are only open-weight. The findings in this space complement our discussion in this paper. We discuss whether transparency aids the security and trustworthiness of LLMs or not. However, independently of this, the transparency claims of models need to be investigated to evaluate whether these models are actually transparent in the first place.

\custompara{Criticism of Open-Source Software Security Claims}
The ``Many-Eyes Principle''~\cite{raymond2001cathedral} proposed that openness could increase the security of software. However, this notion was challenged theoretically~\cite{anderson2002security}, empirically~\cite{Schryen11} (by showing that the number and severity of vulnerabilities in closed- and open-source software do not differ significantly), through examples of vulnerabilities in open-source code that evaded detection for years~\cite{DurumericKAHBLWABPP14}, and through examples of supply-chain attacks~\cite{OhmPS020}. We apply a similar lens to LLMs, analyzing to what extent transparency can help and where it may be harmful. As described in \cref{sec:oss_vs_LLM}, arguments from open-source software cannot be applied to LLMs directly, and thus, we provide an analysis for LLMs based on prior research.
\section{Conclusion}

In this paper, we analyzed the security and trustworthiness of fully transparent LLMs. We conclude that, unlike for open-source software, transparency in LLMs is not effective at removing the need for users to trust the model developer. This mainly comes from two factors. 1) The non-determinism and cost of LLM training make it infeasible for users to verify that the model they are running actually resulted from using the published training pipeline on the published training data. 2) LLMs are too complex (i.e., their training data is too vast to analyze in the required depth, and their weights do not enable robust predictions about the LLM's behavior), making it easier to hide malicious capabilities. While some may argue that transparency alone could disincentivize malicious behavior, it is unclear how true this is if transparency does not help users significantly in establishing trust and security. Furthermore, transparency may reduce the security of LLMs. The generative nature of LLMs means that they can create unsafe outputs even if the training data and pipeline are safe, and thus an adversary may always be able to find a viable attack on the model. Open weights in particular, but also open data and pipelines, increase the attack success rate, since adversaries can use knowledge of the weights, data, and pipeline to improve their attacks. Additionally, patching an LLM in a robust way is costly, making it less likely that the model can be updated quickly as vulnerabilities are detected. Lastly, transparency may increase liability risks, as anyone can search the data for sensitive information, such as copyrighted data or personal information.

This means that transparency for LLMs does not provide the same benefits as open-sourcing does for software, and may even introduce new risks. However, there are also benefits of transparency for LLMs. Transparency allows users to customize the model through finetuning, which is less costly than training a new model from scratch. This allows a user to add specialized capabilities, cultural context, and updated information to the LLM. Furthermore, the user can add their own security hardening to the model by re-aligning the model and adding new security guardrails. Lastly, the user of the model can be sure that they will not lose access to the model as long as they have the hardware to host the model themselves. 

In conclusion, there is no simple guideline (e.g., ``More transparency is always more trustworthy'' or ``Closed models are safer'') that one can follow to decide which level of LLM transparency is best. Instead, it is important to carefully evaluate the LLM developer, the regulatory environment, the use case for the LLM, the type of users the model will have, and the current technology landscape to decide the level of transparency that is best suited. In the following, we highlight two illustrative examples of how such an evaluation might look, and a third example showing where caution is needed. 

\paragraph*{Example I: Users require self-hosting and customization}
Consider the case where the user base of the model wants the freedom to host their model and customize it. In this scenario, the developer must at least provide the weights to enable this type of use. Furthermore, it is reasonable to assume that such a user base will also add new safety guardrails and alignment to the model if instructed to do so, thus limiting some of the safety concerns of transparent models. However, it is less certain whether publishing the training data and pipeline would be beneficial. Currently, if the user wants to verify that the model they received is safe, the training data and pipeline are of limited use. However, publishing the training data could increase litigation risks, and publishing the training pipeline might increase the success rate of attacks on the model. Thus, in this case, an open-weight model release may be judged as best suited. 

\paragraph*{Example II: Users with little technical expertise}
Consider the case where the user base of the model does not have high technology expertise, meaning they are unlikely to adapt the model or host the model themselves. In this case, it is an issue that adversaries may have a higher attack success rate on a transparent model, as users may be more vulnerable even if the model is hosted on a platform with additional safeguards. Furthermore, an open release may provide limited benefits to users as it neither helps them establish trust (neither through their own review of the model, nor through independent audits) nor allows them to take advantage of the additional freedom. Thus, in this case, a closed model release may be judged as best suited. 

\paragraph*{Example III: A nation has strong regulatory requirements}
Consider the case where a nation requires all LLMs to be audited and approved by a government auditor before they can be offered to its citizens. In such a case, a fully transparent release may seem like an obvious choice as a regulator can directly look at all model components. However, this is not the only option. The developer of a closed model could also give selective access (e.g., to the government or an external auditor) to the components of their model, while restricting access to the broader public. They may even allow select parties to do unannounced, periodic checks to ensure that the model they serve is what they promise. While this may not allow the general public to draw their own conclusions, this may be enough for regulatory compliance. In both cases, we note that the government also needs to be careful about what they can actually guarantee, as a government stamp of approval usually inspires trust, which is an issue if unwarranted. They may be able to inspect the weights, data, and pipeline, but if, due to the limited reproducibility of LLMs, they cannot connect the weights to the data and pipeline, then parts of the audit may concern components that may not have actually been used.

\paragraph*{Outlook}
We note that such evaluations need to be continuously updated  as LLMs' capabilities change and the available tools and methods improve. For example, in the future, a new data scanning method may be developed that significantly helps identify backdoors in training data, or, fully attested training may become viable for frontier models. This would make open data significantly more valuable to users, tipping the scale towards open-data releases. Lastly, future research should quantify the increase in security risk of different transparent LLM components (i.e., open weights, open data, open pipeline, and their combinations), so that users and developers can make better risk assessments when deciding what type of model to use or publish.

\bibliographystyle{plainurl}
\bibliography{references}

\end{document}